\documentstyle[12pt,epsf,a4]{article}
\renewcommand{\thefootnote}{\fnsymbol{footnote}}
\begin{document}    

\title{\vskip-3cm{\baselineskip14pt
\centerline{\normalsize BUTP--99/03 \hfill hep-ph/9902480}
\centerline{\normalsize TTP99--10   \hfill February 1999}
}
\vskip.2cm
Non-Singlet Corrections of 
${\cal O}(\alpha_s^2 G_F M_t^2)$ to
$\Gamma(Z\to b\bar{b})$\footnote{This work was supported by DFG under Contract
  Ku 502/8-1 and the {\it Schweizer Nationalfond.}}
\vskip.1cm
}
\author{{K.G. Chetyrkin}$^{a,}$\thanks{Permanent address:
Institute for Nuclear Research, Russian Academy of Sciences,
60th October Anniversary Prospect 7a, Moscow 117312, Russia.}
\,\,and 
{M. Steinhauser}$^b$
  \\[1em]
 {\it (a) Institut f\"ur Theoretische Teilchenphysik,}\\
  {\it Universit\"at Karlsruhe, D-76128 Karlsruhe, Germany}
  \\[.4em]
  {\it (b) Institut f\"ur Theoretische Physik,}\\ 
  {\it Universit\"at Bern, CH-3012 Bern, Switzerland}
}
\date{}
\maketitle

\vspace{-1cm}

\begin{abstract}
\noindent
The partial decay rate of the $Z$ boson into bottom quarks constitutes an
important decay channel. This is mainly due to the virtual presence
of the top quark in the loop diagrams giving rise to correction factors which
are quadratic in the top quark mass.
At one- and two-loop order it turned out that the leading term in the
heavy-top expansion leads to very good approximations to the exact result.
In this work the non-singlet diagrams at ${\cal O}(\alpha_s^2 G_F M_t^2)$ are
considered.
\end{abstract}




\thispagestyle{empty}

\renewcommand{\thefootnote}{\arabic{footnote}}
\setcounter{footnote}{0}

The impressive experimental precision mainly at the Large Electron Positron
collider (LEP) at CERN, the Stanford Linear Collider (SLC) and the FERMILAB
Tevatron in Chicago has made it mandatory to
evaluate higher order quantum corrections to the processes observed in the
experiments~\cite{expthe}.
The strategy to combine experimental information with theoretical 
computations has successfully been applied to the search for the top quark
several years ago. Nowadays the same
concept is used in order to pin down the mass
of the Higgs boson, the only not yet discovered particle of the Standard
Model of elementary particle physics.

An important observable is the decay of the $Z$ boson into bottom.
QCD corrections are known up to ${\cal O}(\alpha_s^3)$ (for a comprehensive
review see~\cite{Rep}).
The electroweak
one-loop corrections are known since quite some time~\cite{AkhBarRie86}.
They have the interesting feature that
the top quark appears virtually in the loop diagrams.
Recently also the full corrections of ${\cal O}(\alpha\alpha_s)$
were completed~\cite{Kat92,CzaKue96,HarSeiSte98}.
The diagrams involving a top quark are considered
in~\cite{HarSeiSte98} where the first five terms in the expansion
for a heavy top quark mass, $M_t$, is computed. It was demonstrated that
these terms approximate the
exact result quite well. Actually it turned out that both at 
${\cal O}(\alpha)$ and ${\cal O}(\alpha\alpha_s)$ a large cancellation
between the sub-leading terms takes place and effectively only the 
leading term proportional to
$G_F M_t^2$~\cite{FleJegRacTar92}
remains. This is a strong motivation
to look at the next order in the strong coupling constant and evaluate the
leading terms.
To the corrections enhanced by the top quark mass only those contributions have
to be considered where a scalar particle, namely the Higgs boson, $H$, the
neutral Goldstone boson, $\chi$, or the charged one, $\phi^\pm$, couples to 
the top quark.
Thus no diagrams have to be considered where the $W$ or $Z$ boson
appear as internal lines.

Corrections of this order were first computed for the
$\rho$ parameter~\cite{Vel77},
the ratio of the charged and neutral current amplitude,
where it turned out that they are quite important~\cite{Avd95}.
Later on also the hadronic Higgs decay was analyzed at ${\cal
  O}(\alpha_s^2G_FM_t^2)$~\cite{CheKniSte96,Ste98}.
In the case of the Higgs boson one can exploit that the scalar coupling is
proportional to the mass which simplifies the construction of an
effective Lagrangian
and especially the subsequent evaluation of the diagrams.
Actually the whole computation could be reduced to the evaluation of two-point
functions. We will see
below that in the case of the $Z$ boson one should also consider
vertex diagrams.

It has become customary to parametrize the corrections proportional to
$M_t^2$ by the quatity
\begin{eqnarray}
X_t &=& \frac{G_F M_t^2}{8\pi^2\sqrt{2}} 
\,,
\end{eqnarray}
respectively the quantity $x_t$ which is defined
using the ${\overline{\rm MS}}$ definition of the top quark mass, $m_t$.

\begin{figure}[b]
  \begin{center}
    \leavevmode
    \epsfxsize=10.cm
    \epsffile[120 620 460 730]{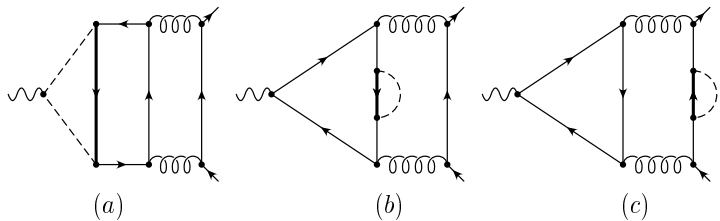}
    \hfill
    \caption[]{\label{fig:sing}
      Singlet diagrams contributing at ${\cal O}(\alpha_s^2 X_t)$
      to the hadronic $Z$ boson decay.
      In $(a)$ and
      $(c)$ the dashed line correspond to the charged Goldstone boson whereas
      in $(b)$ it may also be the Higgs or the neutral Goldstone boson.
      Diagrams $(a)$ and $(b)$ are of universal type whereas $(c)$ constitutes
      a non-universal contribution to $\Gamma(Z\to b\bar{b})$.
      In the displayed examples the thick lines correspond to top quarks
      whereas the thin lines represent bottom quarks.
      }
  \end{center}
\end{figure}
\begin{figure}[t]
  \begin{center}
    \leavevmode
    \epsfxsize=10.cm
    \epsffile[120 460 460 730]{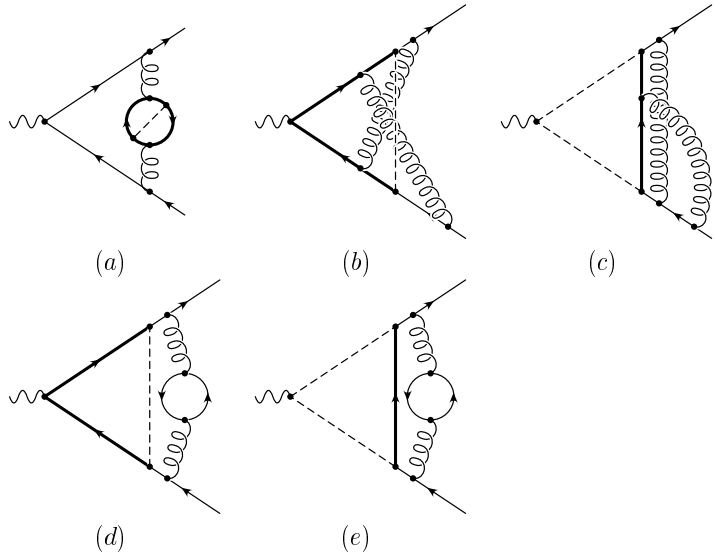}
    \hfill
    \caption[]{\label{fig:nonsing}
      Non-singlet contributions of ${\cal O}(\alpha_s^2 X_t)$
      to the hadronic $Z$ boson decay.
      In $(a)$ the dashed line corresponds to the Higgs boson or the
      neutral or charged Goldstone boson.
      In the diagrams $(b)-(e)$ only the charged Goldstone boson is allowed.
      Diagram $(a)$ is of universal type whereas $(b)-(e)$ constitute
      non-universal contributions to $\Gamma(Z\to b\bar{b})$.
      In the displayed examples the thick lines correspond to top quarks
      whereas the thin lines represent bottom quarks.
      }
  \end{center}
\end{figure}

The quantum corrections to $\Gamma(Z\to b\bar{b})$ are divided into universal
ones which are identical for all quark species and non-universal parts
which are specific for the $Zb\bar{b}$ vertex.
Both the universal and non-universal corrections are divided into singlet and
non-singlet parts. The singlet contributions arise
from those diagrams where the $Z$ boson and the bottom quarks of the final
state couple to different fermion lines.
Another class of singlet contributions is constituted by the diagrams
where the $Z$ boson couples to two
charged Goldstone bosons which in turn form together with two gluons 
a box diagram and the gluons finally couple to the quarks in the final state.
In Fig.~\ref{fig:sing} some sample diagrams are listed.
Fig.~\ref{fig:sing}$(a)$
and $(b)$ are
of universal nature whereas in $(c)$ the Goldstone boson in directly
coupled to the final state bottom quark thus providing a non-universal
contribution.
In this article only non-singlet diagrams will be
computed; the singlet contributions will be considered
elsewhere. 
The universal corrections of ${\cal O}(\alpha_s^2X_t)$ are in part governed
by the $\rho$ parameter~\cite{Avd95}.
A second source for universal corrections 
arise from those diagrams where
only gluons couple to the light quark lines. The gluons split into 
a fermion loop actually formed by bottom and top quarks accompanied by an
additional exchange of a scalar particle (cf. Fig.~\ref{fig:nonsing}$(a)$).
The main focus of this paper is devoted to the evaluation of the non-universal
non-singlet diagrams.
Typical examples are pictured in Fig.~\ref{fig:nonsing}$(b)$-$(e)$.

In a first step an effective Lagrangian is constructed were
the top quark is integrated out. Thereby it is convenient to split the fermion
fields into their left and right part and consider them separately.
It is furthermore necessary to decouple the bottom quark fields using the
relations (see e.g.~\cite{CheKniSte96}):
\begin{eqnarray}
b_{L/R}^{0\prime} &=& \sqrt{\zeta_{2,b}^{0,L/R}} b_{L/R}^{0}
\,,
\end{eqnarray}
where the primes denote the quantities in the effective theory and the
superscript ``0'' reminds that we are still dealing with bare quantities.
The decoupling constants $\zeta_{2,b}^{0,L/R}$ can be computed with the help of
\begin{eqnarray}
\zeta_{2,b}^{0,L/R} &=& 1 + \Sigma_{b,V}^{0h} \mp \Sigma_{b,A}^{0h}
\,,
\end{eqnarray}
where
$\Sigma_{b,V}^{0h}$ and $\Sigma_{b,A}^{0h}$ is the vector and axial-vector
part of the bottom quark self energy.
Here only the hard part, i.e. those diagrams containing the top quark,
has to be computed which is indicated by the index ``h''.
Finally the part of the effective Lagrangian describing the interaction
of the $Z$ boson to bottom quarks has the form
\begin{eqnarray}
{\cal L}_{\rm eff} &\sim&
\left[
  C_L^0 \bar{b}_L^{0\prime} \gamma^\mu b_L^{0\prime}
  +
  C_R^0 \bar{b}_R^{0\prime} \gamma^\mu b_R^{0\prime}
\right] Z_\mu
\,.
\end{eqnarray}
The residual dependence on $M_t$ is contained
in the coefficient functions $C_{L/R}^0$. They are obtained from the hard
part of the $Zb\bar{b}$
vertex:
\begin{eqnarray}
C_{L/R}^0 &=& \frac{\Gamma_{Zb\bar{b}}^{h,L/R}}{\zeta_{2,b}^{0,L/R}}
\,.
\label{eq:coef}
\end{eqnarray}
Here, the left and right part of the $Zb\bar{b}$ vertex are defined through:
\begin{eqnarray}
\Gamma_{Zb\bar{b},\mu}^{h}\,\,=\,\, \gamma_\mu\left[
  \Gamma_{Zb\bar{b}}^{h,V} + \gamma_5 \Gamma_{Zb\bar{b}}^{h,A}
\right]
\,,
&\quad&
\Gamma_{Zb\bar{b}}^{h,L/R}\,\,=\,\,
\Gamma_{Zb\bar{b}}^{h,V} \mp
\Gamma_{Zb\bar{b}}^{h,A}
\,,
\end{eqnarray}
where it is understood that in addition to the top-induced diagrams also the
tree-level terms are included.
The coefficient functions of Eq.~(\ref{eq:coef}) are finite after the coupling
constant $\alpha_s$ and the mass of the top quark are expressed through their
renormalized counterparts. This is because the vector and axial-vector
currents have vanishing anomalous dimension as long as only 
non-singlet diagrams are considered.
Thus from now on the index ``0'' is omitted.

In order to evaluate the partial decay rate of the $Z$ boson into bottom
quarks at ${\cal O}(\alpha_s^2X_t)$ one has to evaluate the coefficient
functions up to this accuracy. Furthermore the pure QCD corrections
in the effective theory are needed up to order $\alpha_s^2$.
It can be taken over from~\cite{QCD5} and reads:
\begin{eqnarray}
\delta^{(5),QCD}
&=&
1 + 
\frac{\alpha_s^{(5)}(\mu)}{\pi} +
\left(\frac{\alpha_s^{(5)}(\mu)}{\pi}\right)^2 
\bigg[\frac{365}{24} - 11\,\zeta_3
\nonumber \\ &&\mbox{} +  n_l
\left( -\frac{11}{12}+\frac{2}{3}\,\zeta_3\right) + 
\left( -\frac{11}{4} + \frac{1}{6} n_l \right) 
 \ln\frac{M_Z^2}{\mu^2} \bigg] 
\,,
\end{eqnarray}
where $n_l=5$ is the number of light quarks and
$\zeta_i$ is Riemann's Zeta function with the value
$\zeta_3\approx1.202056903$.

The computation of the decoupling constants for the bottom quark field up to
order $\alpha_s^2 X_t$ has been performed in~\cite{CheKniSte96}.
The only missing pieces are the vector and axial-vector contributions to
the hard part of the $Zb\bar{b}$ vertex. Some sample diagrams are listed in
Fig.~\ref{fig:nonsing}. 
As mentioned above only those diagrams have to be taken into account which
contain a virtual top quark.
Note that for the very calculation
it is possible to nullify all external momenta.
At one-loop order only two diagrams have to be considered. This increases to 
$19$ at the two-loop level which is still feasible by hand. In the order we
are interested in, however, more than 350 diagrams have to be considered, which
makes the extensive use of computer algebra necessary. 
For the present calculation the package {\tt GEFICOM}~\cite{geficom}
has been used. It passes the generation of the diagrams to
{\tt QGRAF}~\cite{Nog93} and uses for the very computation of the integrals
the program {\tt MATAD}~\cite{Stediss} which is written in
{\tt FORM}~\cite{form}
for the purpose to compute one-, two- and three-loop vacuum graphs.
For a recent review concerned with the automatic computation of Feynman
diagrams see~\cite{HarSte98}.

Expressed in terms of the ${\overline{\rm MS}}$ top quark mass the 
result for the coefficient functions read:
\begin{eqnarray}
C_L &=& 
\frac{e}{s_\theta c_\theta}\Bigg\{
-\frac{1}{2}+\frac{1}{3}s_\theta^2
+x_t\Bigg[
  1
  + \frac{\alpha_s^{(6)}(\mu)}{\pi}\,C_F\,\left(
            2
          - \frac{3}{2}\zeta_2
          + \frac{3}{2}\ln\frac{\mu^2}{m_t^2}
    \right)
\nonumber\\&&\mbox{}
  + \left(\frac{\alpha_s^{(6)}(\mu)}{\pi}\right)^2\Bigg(
         C_F^2 \bigg(
          - \frac{49}{192}
          - \frac{199}{48}\zeta_2
          + \frac{253}{12}\zeta_3
          - \frac{77}{8}\zeta_4
          + \frac{5}{4}B_4
          - \frac{5}{8}D_3
          - \frac{1053}{32}S_2
\nonumber\\&&\mbox{}
          + \left(\frac{15}{16}
              - \frac{9}{4}\zeta_2\right)\ln\frac{\mu^2}{m_t^2}
          + \frac{9}{8}\ln^2\frac{\mu^2}{m_t^2}
          \bigg)
       + C_AC_F  \bigg(
            \frac{461}{64}
          - \frac{99}{32}\zeta_2
          - \frac{187}{24}\zeta_3
          + \frac{61}{16}\zeta_4
\nonumber\\&&\mbox{}
          - \frac{5}{8}B_4
          + \frac{5}{16}D_3
          + \frac{1053}{64}S_2
          + \left(\frac{185}{48}
              - \frac{11}{8}\zeta_2\right)\ln\frac{\mu^2}{m_t^2}
          + \frac{11}{16}\ln^2\frac{\mu^2}{m_t^2}
          \bigg)
\nonumber\\&&\mbox{}
       + C_FTn_l \bigg(
          - \frac{95}{48}
          + \frac{4}{3}\zeta_2
          - \zeta_3
          + \left( - \frac{13}{12}
              + \frac{1}{2}\zeta_2\right)\ln\frac{\mu^2}{m_t^2}
          - \frac{1}{4}\ln^2\frac{\mu^2}{m_t^2}
          \bigg)
\nonumber\\&&\mbox{}
       + C_FT \bigg(
            \frac{149}{240}
          - \frac{1}{60}\zeta_2
          - \frac{35}{8}\zeta_3
          + \frac{729}{40}S_2
          + \left( - \frac{13}{12}
              + \frac{1}{2}\zeta_2\right)\ln\frac{\mu^2}{m_t^2}
          - \frac{1}{4}\ln^2\frac{\mu^2}{m_t^2}
          \bigg)
    \Bigg)
  \Bigg]
\Bigg\}
\nonumber\\&=&\mbox{}
\frac{e}{s_\theta c_\theta}\Bigg\{
-\frac{1}{2}+\frac{1}{3}s_\theta^2
+x_t\Bigg[
  1
  + \frac{\alpha_s^{(6)}(\mu)}{\pi}\left(
            \frac{8}{3}
          - 2\zeta_2
          + 2\ln\frac{\mu^2}{m_t^2}
    \right)
\nonumber\\&&\mbox{}
  + \left(\frac{\alpha_s^{(6)}(\mu)}{\pi}\right)^2\Bigg(
  \frac{62149}{2160} 
- \frac{21337}{1080}\zeta_2 
+ \frac{367}{108}\zeta_3
- \frac{67}{36}\zeta_4
- \frac{5}{18}B_4 
+ \frac{5}{36}D_3
+ \frac{1557}{80}S_2 
\nonumber\\&&\mbox{}
+ n_l\left(
  -\frac{95}{72} 
  +\frac{8}{9}\zeta_2
  -\frac{2}{3}\zeta_3
  \right)
+ \left(
  \frac{589}{36}
  - \frac{55}{6}\zeta_2
  + n_l\left(-\frac{13}{18} + \frac{1}{3}\zeta_2\right)
\right)\ln\frac{\mu^2}{m_t^2}
\nonumber\\&&\mbox{}
+ \left(\frac{55}{12} - \frac{1}{6}n_l\right) \ln^2\frac{\mu^2}{m_t^2}
    \Bigg)
  \Bigg]
\Bigg\}
\,,
\\
C_R &=& 
\frac{e}{s_\theta c_\theta}
\frac{1}{3}s_\theta^2
\nonumber
\,,
\end{eqnarray}
with $m_t=m_t(\mu)$. 
After the second equal sign the colour factors $C_F=4/3, C_A=3$ and
$T=1/2$ have been inserted. $\zeta_2=\pi^2/6$ and $\zeta_4=\pi^4/90$.
$s_\theta$ and $c_\theta$ are the sine and cosine of the weak mixing angle.
The constants $B_4$, $D_3$ and $S_2$ typically appear in the result of
three-loop vacuum integrals and read~\cite{Bro92,Avd95,Bro98}:
\begin{eqnarray}
S_2&=&{4\over9\sqrt3}\mbox{Cl}_2\left({\pi\over3}\right)
\,\,\approx\,\,0.260\,434
\,,
\nonumber\\
D_3&=&6\zeta_3-\frac{15}{4}\zeta_4
     -6\left(\mbox{Cl}_2\left({\pi\over3}\right)\right)^2
\,\,\approx\,\,-3.027\,009
\,,
\nonumber\\
B_4&=&16\mbox{Li}_4\left({1\over2}\right)-{13\over2}\zeta_4-4\zeta_2\ln^22
+{2\over3}\ln^42
\,\,\approx\,\,-1.762\,800
\,.
\end{eqnarray}
Note that according to the QED Ward identity the universal corrections
induced by the diagrams in Fig.~\ref{fig:nonsing}$(a)$
cancel in the coefficient functions
against the corresponding part in the quark self energy.
As we consider in addition the bottom quark to be massless the right-handed
coefficient function sticks to its Born value.
Using the relation between $m_t$ and the on-shell mass
$M_t$~\cite{GraBroGraSch90}
one gets:
\begin{eqnarray}
C_L^{OS} &=& 
\frac{e}{s_\theta c_\theta}\Bigg\{
-\frac{1}{2}+\frac{1}{3}s_\theta^2
+X_t\Bigg[
  1
  -2\zeta_2\frac{\alpha_s^{(6)}(\mu)}{\pi}
  + \left(\frac{\alpha_s^{(6)}(\mu)}{\pi}\right)^2\Bigg(
\frac{1054}{135} 
- \frac{19897}{1080}\zeta_2
+ \frac{403}{108}\zeta_3
\nonumber\\&&\mbox{}
- \frac{67}{36}\zeta_4
- \frac{4}{3}\zeta_2\ln2
- \frac{5}{18} B_4
+ \frac{5}{36} D_3
+ \frac{1557}{80} S_2
 + n_l\left(-\frac{1}{3} 
   + \frac{14}{9}\zeta_2
   - \frac{2}{3}\zeta_3
\right) 
\nonumber\\&&\mbox{}
+ \left(
 -\frac{31}{6} 
 +\frac{1}{3}n_l
\right) \zeta_2 \ln\frac{\mu^2}{M_t^2} 
\Bigg)
  \Bigg]
\Bigg\}
\,,
\\
C_R^{OS} &=& C_R
\nonumber
\,.
\end{eqnarray}

Let us now turn to a brief numerical discussion of the new results.
The decay rate can be computed with the help of
\begin{eqnarray}
\Gamma(Z\to b\bar{b}) &=&
\frac{N_C M_Z}{24\pi} \left(C_L^2 + C_R^2\right)\,\delta^{(5),QCD}
\,.
\label{eq:gamzbb}
\end{eqnarray}
Actually two scales are involved in the process, namely $M_Z$ and the mass of
the top quark. The resummation of potentially large logarithms is, however,
trivial as both $C_{L/R}$ and $\delta^{(5),QCD}$ are separately
renormalization group invariant. Thus the scale parameter $\mu$ may be set to
$m_t$, respectively, $M_t$ in the coefficient functions and to $M_Z$
in the massless corrections.
For these choices the numerical expansions of the ingredients for
Eq.~(\ref{eq:gamzbb}) read:
\begin{eqnarray}
\delta^{(5),QCD} &=&
1 
+\frac{\alpha_s^{(5)}(M_Z)}{\pi}
+1.409\left(\frac{\alpha_s^{(5)}(M_Z)}{\pi}\right)^2 
\,,
\nonumber\\
C_L 
&=&
\frac{e}{s_\theta c_\theta}\Bigg\{
-\frac{1}{2}+\frac{1}{3}s_\theta^2
+x_t\Bigg[
  1
  - 0.623\frac{\alpha_s^{(6)}(\mu_t)}{\pi}
  + 0.190\left(\frac{\alpha_s^{(6)}(\mu_t)}{\pi}\right)^2
\Bigg]
\Bigg\}
\,,
\nonumber\\
C_L^{OS} 
&=&
\frac{e}{s_\theta c_\theta}\Bigg\{
-\frac{1}{2}+\frac{1}{3}s_\theta^2
+X_t\Bigg[
  1
  - 3.290\frac{\alpha_s^{(6)}(M_t)}{\pi}
  - 9.288\left(\frac{\alpha_s^{(6)}(M_t)}{\pi}\right)^2
\Bigg]
\Bigg\}
\,,
\label{eq:cnum}
\end{eqnarray}
with $\mu_t=m_t(m_t)$.
$n_l=5$ has been chosen.

Concerning the enhanced corrections of ${\cal O}(X_t)$ to the coefficient
functions the same observations can be made as for the
$\rho$ parameter~\cite{Avd95}
and the various quantities in connection with the
Higgs decay~\cite{KniSte95,CheKniSte96}:
Expressed in terms of the on-shell top quark mass the leading order
term is ``screened'' by the QCD corrections as they enter with a different
sign. On the other hand,
the coefficients turn out to be much
smaller in the ${\overline{\rm MS}}$ scheme. Actually the coefficient in front
of the three-loop term is smaller by a factor of 50 as compared to the
corresponding one in the on-shell scheme.
Furthermore the sign is alternating which also 
indicates a faster convergence if the ${\overline{\rm MS}}$ mass is used for
the parameterization.

Inserting Eqs.~(\ref{eq:cnum}) into Eq.~(\ref{eq:gamzbb}) finally leads 
to the following $M_t^2$-enhanced terms:
\begin{eqnarray}
\Gamma^{x_t}(Z\to b\bar{b}) &=&
\frac{N_C M_Z \alpha}{6 s_\theta^2 c_\theta^2}
\left(-1+\frac{2}{3}s_\theta^2\right) x_t
\left[ 1 + 0.0161 + 0.0014 \right]
\,,
\label{eq:msnum}
\\
\Gamma^{X_t}_{OS}(Z\to b\bar{b}) &=&
\frac{N_C M_Z \alpha}{6 s_\theta^2 c_\theta^2}
\left(-1+\frac{2}{3}s_\theta^2\right) X_t
\left[ 1 - 0.074 - 0.0130 \right]
\,,
\end{eqnarray}
where the values $\alpha_s^{(5)}(M_Z)=0.118$, $\alpha_s^{(6)}(M_t)=0.107$
and $\alpha_s^{(6)}(m_t)=0.108$ have been used. 
The numbers in the squared brackets correspond to the corrections with
increasing power in $\alpha_s$.
The index $OS$ reminds on the definition of the top quark mass used and as
before the index ${\overline{\rm MS}}$ is suppressed.
In the $\overline{\rm MS}$ scheme the second order QCD corrections amount to
roughly 9\% of the first order ones, however, the overall size is quite
small. The order $\alpha_s$ corrections to the leading $X_t$ term 
in the on-shell scheme is almost by a factor of five larger than in the
$\overline{\rm MS}$ scheme and the corresponding ${\cal O}(\alpha_s^2)$ 
corrections amount to almost 18\% of the ${\cal O}(\alpha_s)$ term.
It is actually almost as large as the ${\cal O}(\alpha_s)$ term in
Eq.~(\ref{eq:msnum}).

To summarize, quantum corrections of ${\cal O}(\alpha_s^2 x_t)$ to the decay
of the $Z$ boson into bottom quarks have been computed.
If we assume that the observations made at one- and two-loop level
are also true at order $\alpha_s^2 X_t$ a substantial part of the corrections
is available.
Expressed in terms of the ${\overline{\rm MS}}$ mass they turn out to be tiny.
In the on-shell scheme the quantum corrections are much larger
and they screen the leading $X_t$ term by almost 9\%.
Note that the newly computed term of ${\cal O}(\alpha_s^2 x_t)$ makes it
possible to use the combination of the three-loop $\rho$
parameter~\cite{Avd95} and
the partial width $\Gamma(Z\to b\bar{b})$ in a consistent way.

\end{document}